\documentclass[12pt,preprint]{aastex}

\def\arcsec{$^{\prime\prime}$}

\def\arcmin{$^{\prime}$}

\def\sol{$_\odot$}
\def\o3hbe{[O~{\sc iii}]1/H$\beta$}
\def\s2hal{[S~{\sc ii}]/H$\alpha$}
\def\n2hal{[N~{\sc ii}]/H$\alpha$}

\def\hal{H$\alpha$}

\def\neii{[Ne~{\sc ii}]}
\def\neiii{[Ne~{\sc iii}]}
\def\microns{$\mu$m}

\def\siii{[S~{\sc iii}]}
\def\deg{$^{\circ}$}

\def\vs1{\vspace*{0.25in}}
\def\vs1m{\vspace*{-0.10in}}
\def\hs1{\hspace*{0.25in}}
\def\aj{AJ}                     
               
\def\apj{ApJ}
\def\apjl{ApJ}
\def\apjs{ApJS}

\def\aap{A\&A}

\def\mnras{MNRAS}

\shorttitle{IRS spectral mapping of NGC 253}
\shortauthors{Devost et al. (2004)}

\begin{document}

\title{Spitzer-IRS mapping of the inner kpc of NGC 253: Spatial distribution of the \neiii, PAH 11.3 \microns\ and H$_2$ (0-0) S(1) lines and a gradient in the \neiii/\neii\ line ratio.}

\author{Daniel Devost\altaffilmark{1},
Bernhard R. Brandl\altaffilmark{1},\altaffilmark{2},
L. Armus\altaffilmark{3},
D. J. Barry\altaffilmark{1},
G.C. Sloan\altaffilmark{1},
Vassilis Charmandaris\altaffilmark{1},
Henrik Spoon\altaffilmark{1},
Jeronimo Bernard-Salas\altaffilmark{1},
and James R. Houck\altaffilmark{1}}

\email{devost@isc.astro.cornell.edu}

\altaffiltext{1}{Center for Radiophysics and Space Research, Cornell University,
219 Space Sciences Building, Ithaca, NY 14853-6801, USA}
\altaffiltext{2}{Sterrewacht Leiden, PO Box 9513 2300 RA Leiden, The Netherlands}
\altaffiltext{3}{{\em Spitzer} Science Center, MS 220-06, California Institute of Technology, Pasadena, CA, 91125}

\begin{abstract}
  
  We present our early results of the mapping of the nucleus of the
  starburst galaxy NGC~253 and its immediate surroundings using the
  Infrared Spectrograph~\footnote{The IRS was a collaborative venture
    between Cornell University and Ball Aerospace Corporation funded
    by NASA through the Jet Propulsion Laboratory and the Ames
    Research Center.} onboard the Spitzer Space Telescope.  The map is
  centered on the nucleus of the galaxy and spans the inner 800
  $\times$ 688 pc$^2$.  The spatial distribution of the \neiii\ line
  at 15.55~\microns\ and the polycyclic aromatic hydrocarbons feature
  at 11.3 \microns\ peaks at the center while the purely rotationnal
  transition of molecular hydrogen at 17.03~\microns\ is strong over
  several slit positions. We perform a brief investigation of the
  implications of these measurement on the properties of the star
  formation in this region using theories developped to explain the
  deficiency of massive stars in starbursts.

\end{abstract}

\keywords{Molecular hydrogen, Galaxies, Starburst, Ionization, Massive stars, Starburst Galaxies}

\section{Introduction}

NGC~253 is a well studied nearby starburst galaxy located at a
distance of 2.5 Mpc \citep{pence80} and is mainly powered by an
episode of central star formation. This episode is generally believed
to be caused by material brought to the center of the galaxy by a bar
which is clearly visible in the 2MASS image \citep{2mass}.  The
presence of a bar from both morphological and kinematical evidence,
and its role as starburst trigger has been extensively discussed in
the literature (Das, Anantharamaiah, \& Yun 2001 and references
therein, Scoville et al. 1985, Engelbracht et al. 1998).

Four luminous super star clusters were discovered in the central
region by \cite{watsonetal96} with the {\em Hubble Space Telescope}.
They derived a bolometric luminosity for the brightest cluster of 1.3
$\times$ 10$^9$L\sol\ which corresponds to 10\% of the luminosity of
within a region of a radius of 15\arcsec\ at the nucleus. Also,
observations of \hal\ emission and earlier {\em Einstein} and {\em
  ROSAT} x-ray data \citep[see][and references therein]{ptaketal97}
revealed a starburst-driven wind, emanating from the nucleus along the
minor axis of the galaxy. This wind was well delineated by {\em
  Chandra} \citep{stricklandetal00}.  \cite{engelbrachtetal98} studied
NGC 253 using near-infrared and mid-infrared lines and concluded that
the properties of the Initial Mass Function (IMF) in the starbursting
region is similar to a Miller-Scalo IMF that has a deficiency in low
mass stars. Also, \cite{boekeretal98} and \cite{ketoetal99} presented
maps of \neii\ 12.81 \microns\ and PAH emission.

Observations in the mid-infrared from space by the {\em Infrared Space
  Observatory} (ISO) were used to identify fine structure lines,
molecular hydrogen lines and emission from polycyclic aromatic
hydrocarbons (PAH)
\citep{strumetal00,rigopoulouetal02,forsteretal03,vermaetal03}.  These
authors mainly studied the nuclear region of the galaxy with the Short
Wavelength Spectrograph (SWS) and ISOCAM.  \cite{rigopoulouetal02}
detected several mid-infrared H$_2$ lines in the SWS spectrum of
NGC~253. They derived an H$_2$~(0--0)~S(2)/S(1) temperature of 380~K
with aperture sizes of 14\arcsec$\times$20\arcsec\ and
14\arcsec$\times$27\arcsec\ for S(2) and S(1) respectively.
\citet{thornleyetal00} measured the \neiii/\neii\ line ratio using SWS
and found a value of 0.06. Since the Ne atom and Ne$^+$ ion have an
ionisation potential of 21.56 and 40.95 eV respectively, this ratio
traces the massive star content because it is very sensitive to the
spectral shape of the UV radiation field.  \citet{thornleyetal00}
addressed specifically this issue by modeling the neon ratio as a
function of the upper mass cutoff. They used evolutionnary synthesis
code an photoionisation code to predict the neon line ratio around a
cluster of stars.  Their results show that a \neiii/\neii\ line ratio
of 0.06 is consistent with an upper mass cutoff around 25 M\sol.
However, they found that for short-lived bursts (1 Myr), the measured
neon line ratio is consistent with the formation of massive (50$-$100
M\sol) stars. \cite{vermaetal03} also analyzed the SWS fine-structure
line mainly in the context of gas-phase abundances.  Their neon line
fluxes indicate a slightly higher \neiii/\neii\ line ratio of 0.08.
They found a correlation between the neon ratio and metallicity among
the SWS starburst sample but excitation for a given metallicity was
relatively low compared to local H II regions.  This problem with the
deficiency of massive stars in starbursts was also found by
\citet{puxleyetal89} and \citet{doyonetal94}.

In this letter, we present spectral maps of \neiii, H$_2$~(0--0)~S(1)
and the PAH feature at 11.3 \microns. The sensitivity and small
aperture size of the IRS gives us the opportunity to do spatial mapping
over small scales of the mid-infrared properties of the central region of
NGC~253. A total of 45 positions were observed around the bright
starburst seen at the center of the galaxy. These observations allowed
us to build the map of the \neiii/\neii\ line ratio where a spatial
gradient is detected. The observations and data reduction are
presented in \S~\ref{obs} and the maps are presented in
\S~\ref{analysis}. The spatial differences in the different maps are
briefly discussed in \S~\ref{discussion}.

\section{Observations and data reduction} \label{obs}

The observations were carried out during the science verification
phase of the Infrared Spectrograph (IRS) onboard the Spitzer Space
Telescope \citep{sirtf} using the Short-High (SH) module
\citep{houcketal04}. The high sensivity and narrow slit sizes of the
IRS were used to map of the central region of NGC~253. The spectra
ranges from 10 to 20~\microns\ with a resolution $R\approx600$.
Although the observations of NGC~253 were non-contiguous on the sky,
we have produced a "sparsely sampled map" to trace the changes in warm
gas and dust properties across the inner starburst.  The map is
centered at a position of RA~$=$~00$^h$47$^m$33.2$^s$
DEC~$=$~-25\deg17\arcmin19\arcsec~(J2000) and the step between each
pointing position is 15\arcsec\ along the length of the slit and
9\arcsec\ along its width.

The map, which consists of 45 different pointings, is
centered on the nucleus of the galaxy and spans an area of
70\farcs8 $\times$ 60\farcs8 at a position angle of 271 degrees.
This corresponds to the central 800~$\times$~688~pc$^2$ of the galaxy at
the assumed distance of 2.5~Mpc. Each slit covers an area of
11\farcs8 $\times$ 5\farcs3 (133~$\times$~60~pc). The exposure time for
each slit position was 88 seconds.

The basic data reduction was performed by the standard IRS pipeline
version~9.5 at the Spitzer Science Center (SSC). The pipeline removes
detector artifacts and cosmic ray signatures and applies the dark and
flatfield corrections (Spitzer Observer's
Manual\footnote{http://ssc.spitzer.caltech.edu/documents/som/},
Chapter 7). Standard full-slit spectral extraction and flux
calibration were performed as described in the SOM. The uncertainty in
the line fluxes is dominated by the absolute flux calibration.  At the
present state, these uncertainties are on the order of 30\%.
Uncertainties on the absolute fluxes have little impact on the
uncertainties of the relative fluxes. Line ratios are mostly
affected by the determination of the continuum near the emission line.
In our dataset, this source of error is below 5\%.

\clearpage
\begin{figure}
\plotone{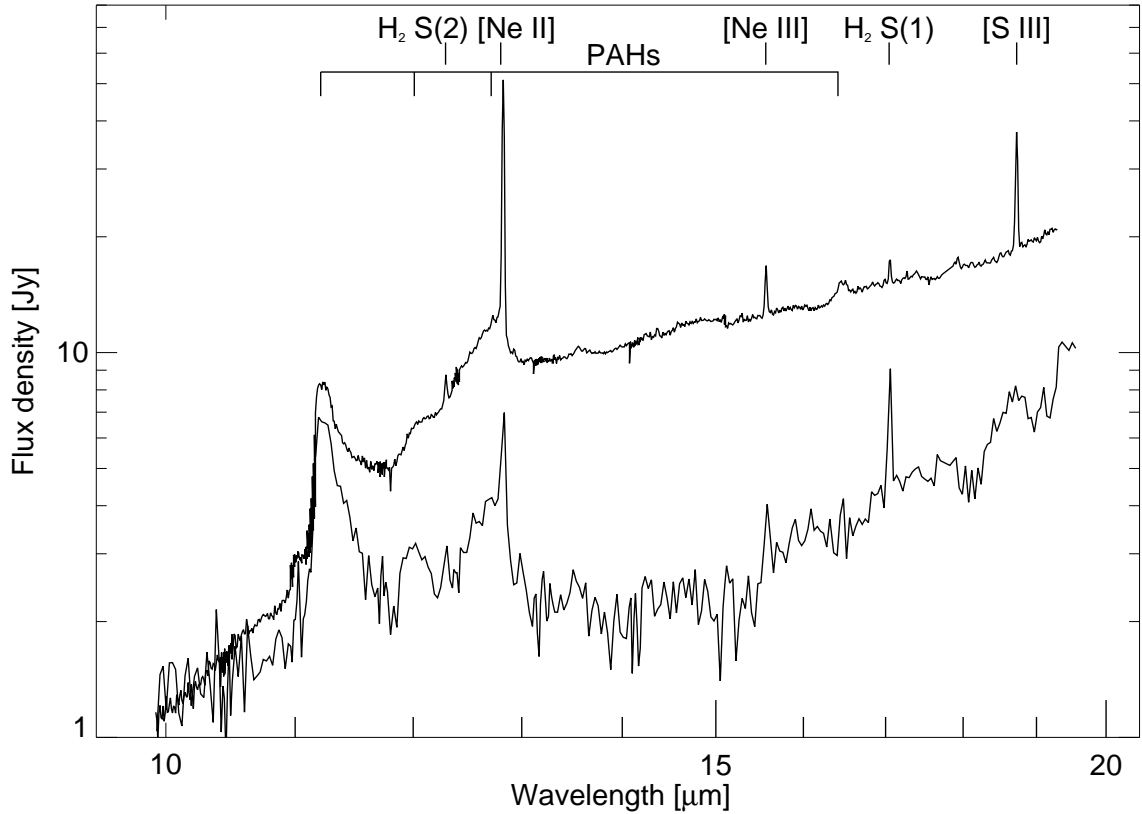}
\caption{The upper line is the spectrum of the central position of the IRS spectral map which
corresponds to the location of the super star clusters of
\cite{watsonetal96}. One of the spectrum off-center (lower line) is shown for comparison.
The units on this spectrum are scaled up by a factor 40.
\label{speccenter}} 
\end{figure}
\clearpage

The 12.8~\microns\ \neii, 15.55~\microns\ \neiii\ and the
17.03~\microns\ H$_2$ (0--0)~S(1) lines are detected with a signal to
noise ($S/N$) higher than 3, our detection limit, on all the spectra.
Line fluxes were measured by fitting a gaussian profile after
substracting a second order baseline fit. Table~\ref{fluxtable}
presents the values of the line fluxes for the central position of the
map, which corresponds to the nuclear region of the galaxy.  Figure
\ref{speccenter} shows the spectrum at the central position and a
typical off-center spectrum. The values measured for the central
spectrum are shown in Table \ref{fluxtable}.
\clearpage
\begin{deluxetable}{lcc}
\tabletypesize{\scriptsize}
\tablecaption{Line fluxes at the nuclear position of NGC 253 \label{fluxtable}}
\tablewidth{0pt}
\tablehead{
\colhead{Line} & \colhead{Rest wavelength (\microns)} & \colhead{Total Flux(10$^{-20}$ W cm$^{-2}$)}
}
\startdata
PAH & 11.3 & 228$\pm$68 \\
PAH & 12.0 & 20$\pm$6 \\
H2 0-0 S(2) & 12.27 & 5$\pm$1 \\
PAH & 12.7 & 159$\pm$48 \\
\neii  & 12.81 & 147$\pm$44 \\
\neiii & 15.55 & 12$\pm$3 \\
PAH & 16.5 & 19$\pm$6 \\
H2 0-0 S(1) & 17.03 & 6$\pm$2 \\
\siii & 18.71 & 40$\pm$12 \\
\enddata
\end{deluxetable}
\clearpage
The spectral maps are shown on Figures~\ref{neiii} to \ref{neratio}. The
total value of the flux in a given line is shown as the value for the
whole slit. This way, each staring observation becomes a pixel and the
result is a $5\times9$ sparse map. All the images are overlaid on the
contours of the 2MASS K-band image.

\section{Analysis}\label{analysis}

\begin{figure}
\plotone{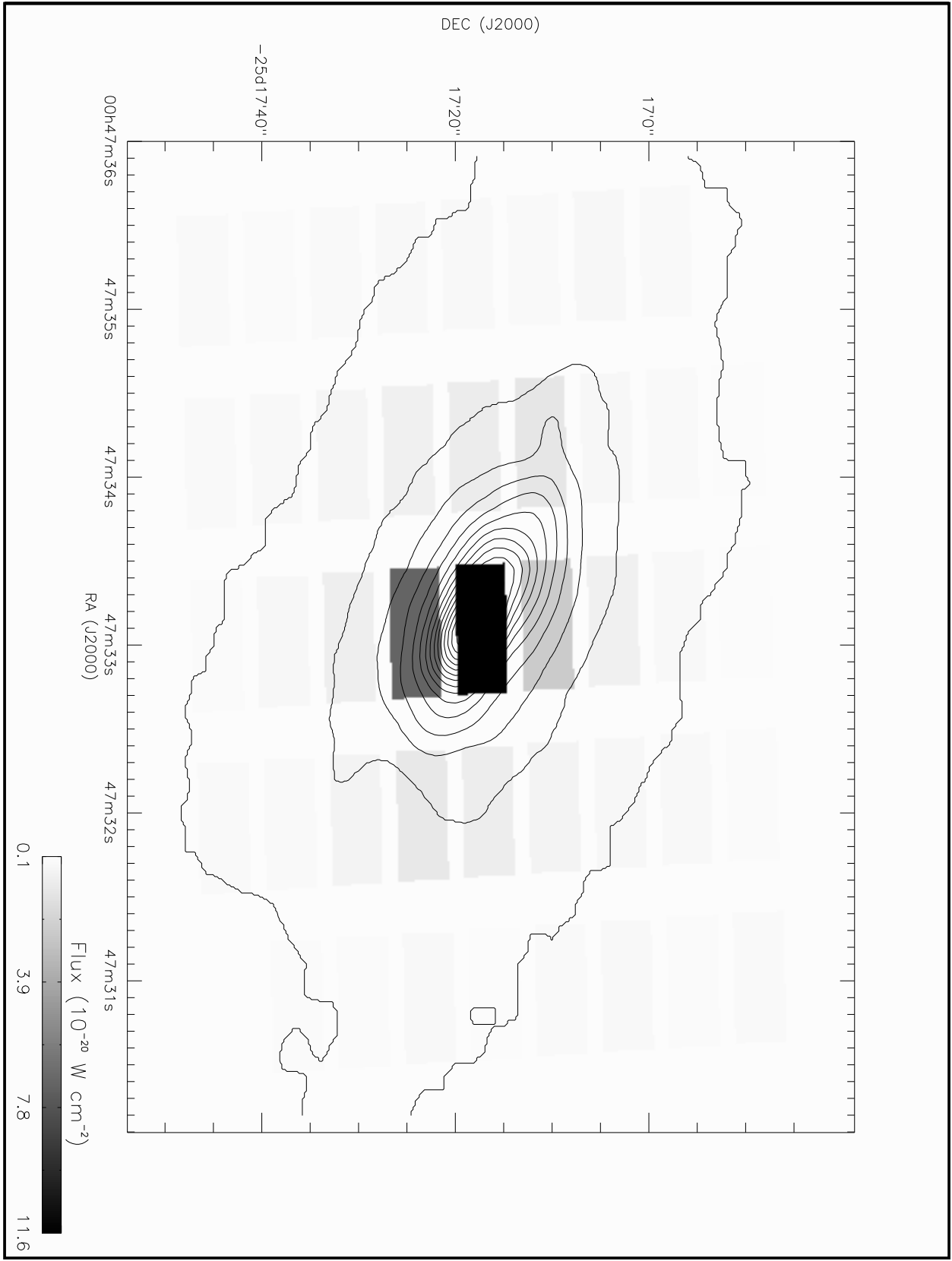}
\caption{Spectral map of Ne {\sc iii} overlaid on the K band contours from the 2MASS image.
The orientation of this map and all the subsequent maps is equatorial. The position
angle of the major axis seen in the 2MASS image is 51\deg\ \citep{pence81}. \label{neiii}}
\end{figure}

\begin{figure}
\plotone{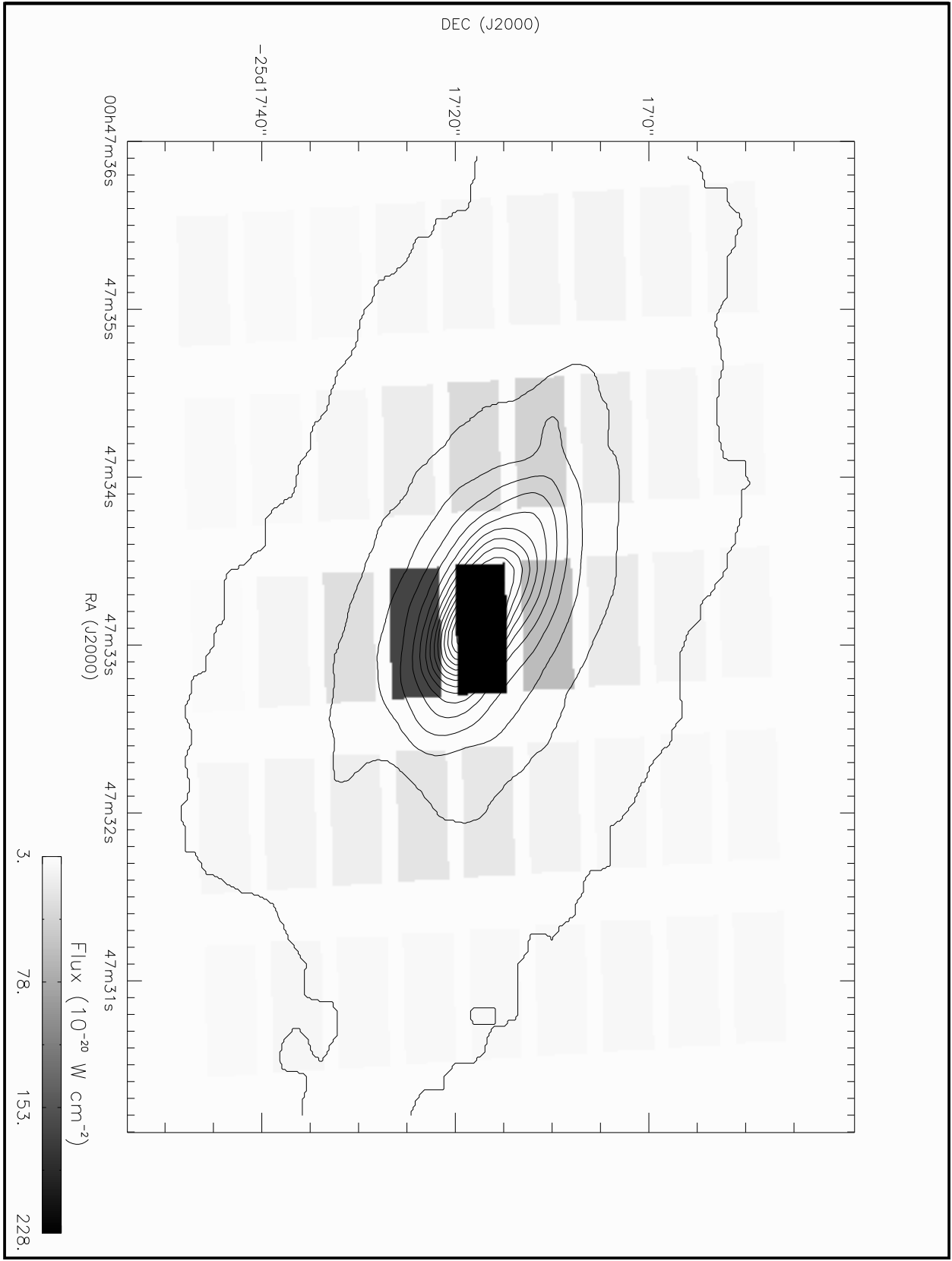}
\caption{\label{pah11}Spectral map of the PAH feature at 11.3 \microns.}
\end{figure}

\begin{figure}
\plotone{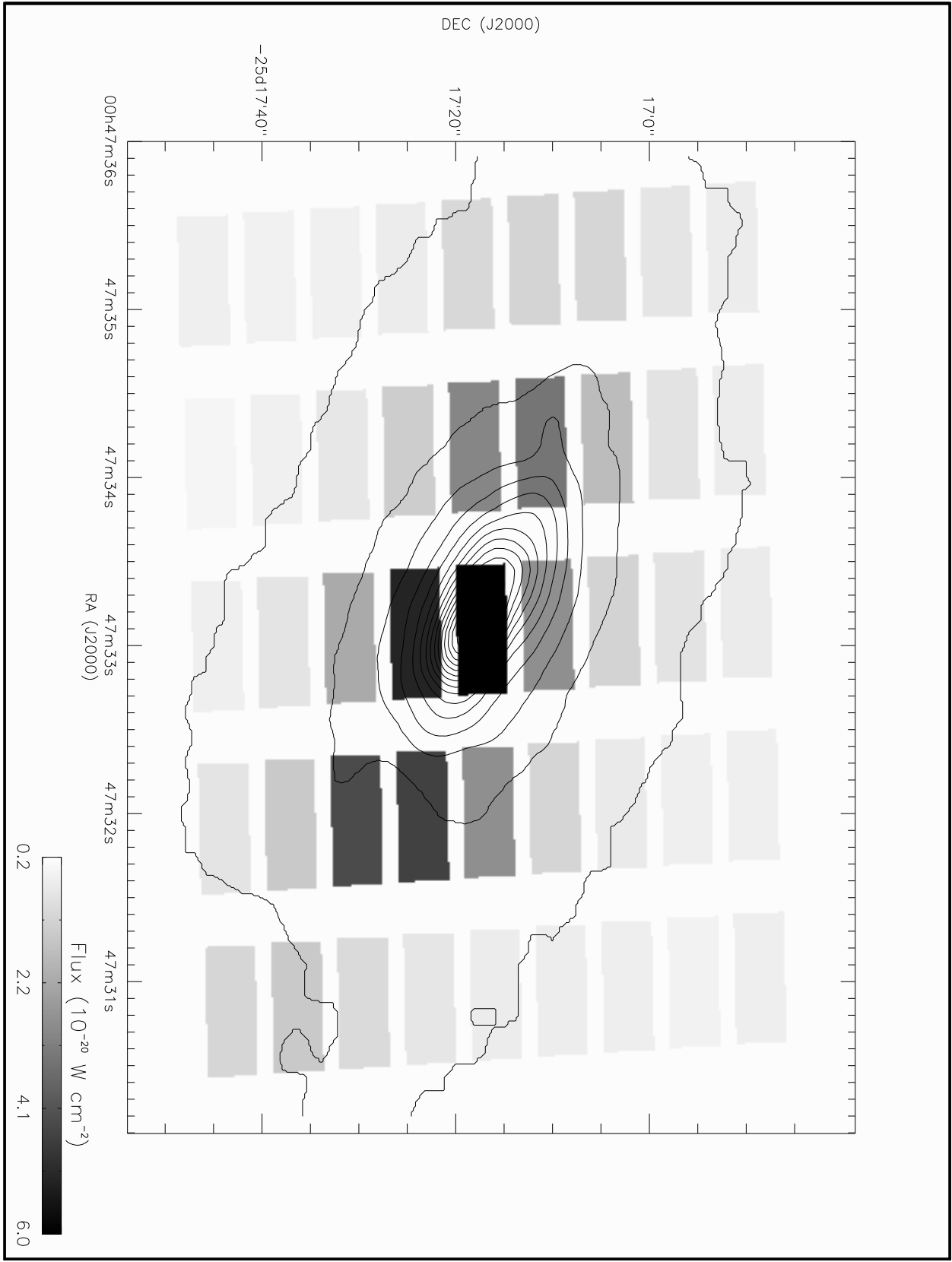}
\caption{Spectral map of H2 (0-0) S(1) over the 2MASS K band contours. \label{h2s1}}
\end{figure}

The \neiii\ map (Figure~\ref{neiii}) shows that the peak emission from
this line is located at the center of the galaxy. Further out, the
absolute line flux drops quickly. The same behavior is also seen with
the PAH 11.3 \microns\ feature (Figure \ref{pah11}). However, the intensity of
the H$_2$ S(1) line is spatially distributed quite differently. We can see
from Figure~\ref{h2s1} that the overall intensity of the H$_2$
(0--0)~S(1) line follows the shape of the major axis of the emission
seen in the 2MASS image.

\clearpage

\clearpage
\clearpage

\clearpage

\begin{figure}
\plotone{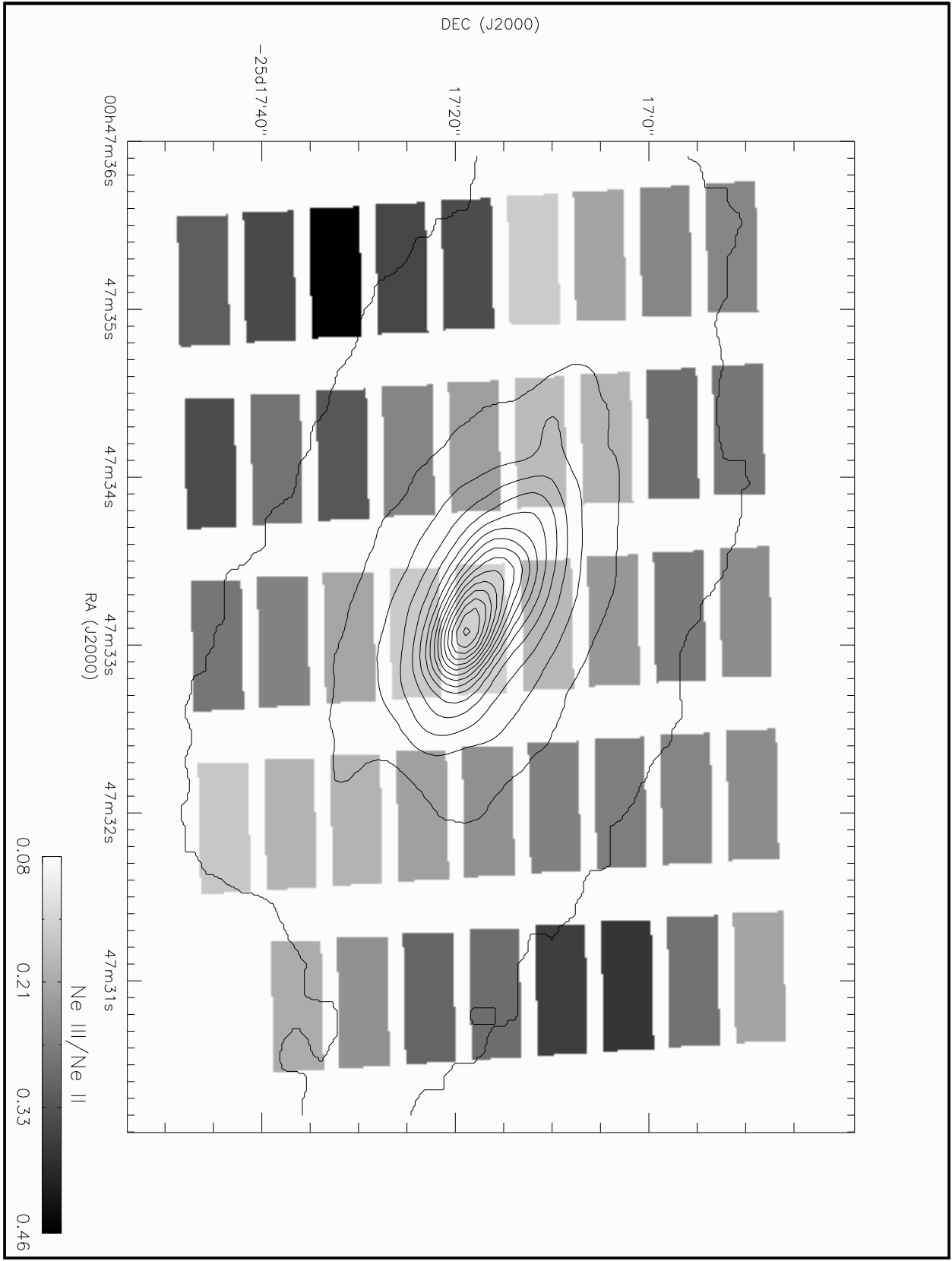}
\caption{Spectral map of the \neiii/\neii line ratio over the 2MASS K band contours. \label{neratio}}
\end{figure}

Figure~\ref{neratio} shows that the \neiii/\neii\ line ratio increases
by a factor 4--6 from the center of the map toward the edges. This
increase is higher than our uncertainties on the relative fluxes (see
section \ref{obs}).  The line ratio at the central pixel is 0.08, in
agreement with what was found with ISO-SWS \citep{vermaetal03}. The
ratio peaks at 0.46 on the south-east edge of the map.

\clearpage
\clearpage

%The H$_2$ temperature map in Figure~\ref{tmap} was derived from the
%H$_2$ 0--0 S(2)/S(1) line ratio using the models of
%\cite{burtonetal92}.  Only spectra with a S/N ratio higher than 3 in
%the S(2) line were incorporated into the map presented in Figure \ref{tmap}.

%\clearpage
%\begin{figure}
%\plotone{f6.ps}
%\caption{Temperature map derived using the H2 (0-0) S(2)/S(1) line ratio. \label{tmap}}
%\end{figure}
%\clearpage

\section{Discussion}\label{discussion}

\subsection{Line fluxes}

The \neiii\ map shows that the strong neon emission is
concentrated in the nucleus and decreases sharply outside that region.
This behavior is also seen in the \neii\ line (not shown here).
It is likely that the super star clusters found by \cite{watsonetal96}
are responsible for the major fraction of the high \neiii\ flux. The
PAH features also follow this trend. These are thought to be excited
in star-forming environments where PAH emission arises predominantly
in photodissociation regions at the interface between ionized and molecular gas
\citep{verstraeteetal96,cesarskyetal96,creteetal99}.

The map of Figure~\ref{h2s1} shows a clear correlation between the
H$_2$~(0--0)~S(1) line and the major axis of the emission seen in
the 2MASS image. This correlation is also seen in the H$_2$
(0--0)~S(2) line (not shown here) and indicates that an excitation
source is located along the major axis and rises the H$_2$ temperature
enough to sustain an elevated emission of the 17.03~\microns\ and the
12.27 \microns\ lines. The purely rotational transitions of molecular hydrogen
can be triggered either by radiation or shocks \citep{burtonetal92}.
In dense ($n \ge 10^5$~cm$^{-3}$) photodissociation regions, UV
photons from massive stars can produce enough energy to raise the
temperature of the H$_2$ gas so that collisional excitation lines are
seen. The molecular gas can also be excited by x-rays
\citep{rigopoulouetal02} or cosmic rays \citep{bradfordetal03}.
The fact that the H$_2$ S(1) line and the PAH 11.3 \microns\ line
don't trace each other may be indicative that different mechanism
maybe at work at the central position and along the major axis.

%\subsection{Hydrogen temperature}
%
%The temperature map of the warm H$_2$ component derived from the
%S(2)/S(1) line ratio suggest that the central value of the temperature
%is $\sim$460~K.  Around the nuclear positions, the H$_2$ temperature drops
%to values of approximately 265~K.  Note that the value derived by
%\cite{rigopoulouetal02} is somewhere between these two values,
%consistent with the different aperture sizes in the measurements. 

\subsection{Spatial variation of the ionization field} \label{spdiff}

The observed increase of the \neiii/\neii\ line ratio (Figure
\ref{neratio}) implies a harder ionization field outside of the
nuclear region of NGC~253. To first order, this means that the
radiation field outside the nucleus is harder than away from the
nucleus. In this section, we breifly review the various physical
conditions that may lead to an increase of the value of the neon line
ratio. We discuss the impact of a metallicity gradient, a modified
IMF, and an age gradient on the physical properties of the nuclear
region of NGC~253.

\cite{martin-hernandez02} showed that decreasing metallicity leads to
a hardening of the ionization field since it affects the line
blanketing and the characteristics of the stellar wind. However, a
large gradient is needed to account for the observed change. The data
of \cite{martin-hernandez02} show that in order to triple the neon
line ratio, a change in metallicity close to 1~Z\sol\ is required. The
maps presented here only cover a radius of 0.5~kpc so the implied metallicity
gradient needed to reproduce the data is on the order of 2 Z\sol kpc$^{-1}$.

Several authors
\citep{puxleyetal89,doyonetal94,engelbrachtetal98,dohertyetal95,achtermannlacy95,becketal97}
proposed a truncated or steeper IMF to explain the deficiency of
massive stars in starburst galaxies. This would imply that the
properties of the IMF would change from the nucleus to the outer disk.
However, the gradient would imply variations in the IMF only if the stellar
populations in the different regions were nearly the same ages.

The analysis of \cite{thornleyetal00} showed that ageing can cause a
softening of the ionization field where a starburst occurs.  They
pointed out that the observed properties of short-lived starbursts are
very sensitive to the massive star content within the first 10 million
years of a star formation episode. This means that the hardness of the
radiation field, and thus the \neiii/\neii\ ratio, should decay rapidly
if the starburst episode responsible for the ionisation field is
short-lived. Using this model, our results would indicate that the
outside regions are, on average, younger than the nuclear region,
consistent with star formation propagating outward.

\section{Conclusions}

We present spectral line maps of \neiii, H$_2$ (0--0)~S(1) ,
PAH 11.3 \microns\ feature as well as the \neiii/\neii\
ratio and the warm molecular hydrogen temperature of the central
region of NGC~253 and its immediate surroundings.  The strength of the
H$_2$ lines follows the morphology of the major axis while the \neiii\ line
and the PAH 11.3 \microns\ feature peak at the location of the nucleus and
show no strong correlation with the major axis, suggesting
different physical mechanism in and outside the nuclear region
for the PAHs and neon on one side and the molecular hydrogen on the other.

A significant gradient in the \neiii/\neii\ ratio is detected.  The
ratio rises toward the edges of the map with the lowest value being at
the location of the nucleus. To first order, this indicates that the
radiation field is harder outside the nucleus which in turn suggests
that there is a relative deficiency of massive stars in the nucleus of
the galaxy. With our dataset, we were able to briefly explore the
variation of the ionisation field within a {\em single} starburst
galaxy and compare it with models that are usually compared to the
ionisation fields of a {\em sample} of starburst galaxies.  The
spectral-mapping capability of the IRS allows detailed studies of the
local environment of powerful starbursts and provide unprecedented
insight into the star formation process in starbursts.  We presented
here a small sample of our dataset. A more detailled study on the
impact of our observations on the physical properties of the central
region of NGC~253 and the stellar population is in preparation.

\acknowledgments

This work is based on observations made with the Spitzer
Space Telescope, which is operated by the Jet Propulsion Laboratory,
California Institute of Technology under NASA contract 1407. Support
for this work was provided by NASA through Contract Number 1257184
issued by JPL/Caltech.

We would alos like to thank the anonymous referee whose suggestions
and comments helped to improve the paper.

%\appendix

%\section{Appendicial material}

 \end{document}